\documentclass[twocolumn,aps,prl,superscriptaddress,showpacs,secnumroman,showkeys]{revtex4}
\usepackage{amsmath,bm,CJK}%
\usepackage{graphicx}%
\usepackage{color}%
\begin{document}
\begin{CJK*}{GBK}{}
\title{
Primary Isotope Yields and Characteristic Properties of the Fragmenting Source in Heavy-ion Reactions near the Fermi Energies
}
\author{X. Liu}
\email[E-mail at:]{liuxingquan@impcas.ac.cn}
\affiliation{Institute of Modern Physics, Chinese Academy of Sciences, Lanzhou, 730000, China}
\affiliation{University of Chinese Academy of Sciences, Beijing 100049, China}
\author{W. Lin}
\affiliation{Institute of Modern Physics, Chinese Academy of Sciences, Lanzhou, 730000, China}
\affiliation{University of Chinese Academy of Sciences, Beijing 100049, China}
\author{R. Wada}
\email[E-mail at:]{wada@comp.tamu.edu}
\affiliation{Institute of Modern Physics, Chinese Academy of Sciences, Lanzhou, 730000, China}
\author{M. Huang}
\affiliation{Institute of Modern Physics, Chinese Academy of Sciences, Lanzhou, 730000, China}
\author{Z. Chen}
\affiliation{Institute of Modern Physics, Chinese Academy of Sciences, Lanzhou, 730000, China}
\author{G. Q. Xiao}
\affiliation{Institute of Modern Physics, Chinese Academy of Sciences, Lanzhou, 730000, China}
\author{S. Zhang}
\affiliation{Institute of Modern Physics, Chinese Academy of Sciences, Lanzhou, 730000, China}
\affiliation{University of Chinese Academy of Sciences, Beijing 100049, China}
\author{X. Jin}
\affiliation{Institute of Modern Physics, Chinese Academy of Sciences, Lanzhou, 730000, China}
\affiliation{University of Chinese Academy of Sciences, Beijing 100049, China}
\author{R. Han}
\affiliation{Institute of Modern Physics, Chinese Academy of Sciences, Lanzhou, 730000, China}
\author{J. Liu}
\affiliation{Institute of Modern Physics, Chinese Academy of Sciences, Lanzhou, 730000, China}
\author{F. Shi}
\affiliation{Institute of Modern Physics, Chinese Academy of Sciences, Lanzhou, 730000, China}
\author{H. Zheng}
\affiliation{Cyclotron Institute, Texas A$\&$M University, College Station, Texas 77843}
\affiliation{Physics Department, Texas A$\&$M University, College Station, Texas 77843}
\author{J. B. Natowitz}
\affiliation{Cyclotron Institute, Texas A$\&$M University, College Station, Texas 77843}
\author{A. Bonasera}
\affiliation{Cyclotron Institute, Texas A$\&$M University, College Station, Texas 77843}
\affiliation{Laboratori Nazionali del Sud, INFN,via Santa Sofia, 62, 95123 Catania, Italy}
\date{\today}

\begin{abstract}
For central collisions of $^{40}$Ca $+ ^{40}$Ca at 35 MeV/nucleon, the density and temperature of a fragmenting source have been  evaluated in a self-consistent manner using the ratio of the symmetry energy coefficient relative to the temperature, $a_{sym}/T$, extracted from the yields of primary isotopes produced in  antisymmetrized molecular dynamics (AMD) simulations.  The $a_{sym}/T$ values are extracted from all isotope yields using an improved method based on the Modified Fisher Model (MFM). The values of $a_{sym}/T$ obtained, using different interactions with different density dependencies of the symmetry energy term, are correlated with the values of the symmetry energies at the density of fragment formation. Using this correlation, the fragment formation density is found to be $\rho/\rho_0 = 0.67 \pm 0.02$. Using the input symmetry energy value  for each interaction temperature values are extracted as a function of isotope mass $A$. The extracted temperature values are compared with those evaluated from the fluctuation thermometer with a radial flow correction.
\end{abstract}



\maketitle
\end{CJK*}

\section*{I. Introduction}

A wealth of nuclear phenomena, including dynamical processes in nuclear reactions, nuclear structure and nuclear astrophysics are intimately affected by the density dependence of the symmetry energy term in the nuclear equation of state (EOS)~\cite{Baoan2008}. Investigations of the density dependence of the symmetry energy  have been conducted using many observables such as isotopic yield ratios~\cite{Tsang2001}, isospin diffusion~\cite{Tsang2004}, neutron-proton emission ratios~\cite{Famiano2006}, giant monopole resonances~\cite{Li2007}, pygmy dipole resonances~\cite{Klimkiewicz2007}, giant dipole resonances~\cite{Trippa2008}, collective flow~\cite{zak2012} and isoscaling~\cite{Xu2000,Tsang2001_1,Huang2010}. The ratio of the symmetry energy coefficient relative to the temperature, $a_{sym}/T$, extracted from isotope yield ratios~\cite{Huang2010_1, Huang2010_2}, has been studied within the frame work of the Modified Fisher Model (MFM)~\cite{Fisher1967,Minich1982,Hirsch1984,Bonasera2008}.

While temperature is one of the key variables in characterizing    nuclear reactions, it is very difficult to determine the temperature of  hot nuclear matter in a dynamical process. Several nuclear thermometers have been proposed. These  include the slope of energy spectra~\cite{Westfall1982,Jacak1983}, momentum fluctuations~\cite{Wuenschel2010}, double isotope yield ratios~\cite{Albergo1985} and excited state distributions~\cite{Morrissey1984} among others. However they may not be generally applicable in all circumstances and
even for a given system the extracted temperature values from these thermometers may be quite different from each other~\cite{Guo2013}. In our recent work, the isotopic yield ratio method was applied to extract $a_{sym}/T$ values from  the experimentally reconstructed primary fragment yields. These ratios were  compared to those calculated from AMD primary generated fragment yields obtained using Gogny interactions with different density dependencies of the symmetry energy~\cite{Lin2014}. In the analysis, we found that the extracted $a_{sym}/T$ values change according to the interactions used. From the dependence on interaction, the density, symmetry energy and temperature at the time of fragment formation were determined in a self-consistent manner.

In order to perform a more detailed investigation of the relationship between the experimentally extracted observables and input parameters such as the density dependent symmetry energy term in the model, we apply a similar procedure to AMD primary events, using AMD as an event generator. There are three major reasons to use AMD as the event generator for this work.
One is its capability to reproduce the experimental isotope yields. AMD results, such as multiplicity, angular distribution and energy spectra, have often been compared with those from the experimental data for intermediate energy heavy ion collisions and reproduce them reasonably well~\cite{Ono96,Ono99,Ono02,Ono04,Wada98,Wada00,Wada04,Hudan06}. In one of our recent works in Ref.\cite{Lin2014}, the yields of the experimentally reconstructed primary hot isotopes are well reproduced by those of the AMD simulations. Second is to eliminate the secondary cooling effect. As shown in Ref.\cite{Huang2010_1,Lin2014}, the sequential decay of the primary hot isotopes significantly alters the yield distribution and distorts the information inherent in  the primary hot fragment yields. Third is to simplify the initial conditions,  using zero impact parameter to eliminate effects of  transverse flow and neck emission among others. The AMD events are generated for central collisions ($b=0$ fm) of $^{40}Ca + ^{40}Ca$ at 35 MeV/nucleon, using  interactions  having different density dependencies of the symmetry energy term, i.e.,  the standard Gogny interaction which has an asymptotic soft symmetry energy (g0), the Gogny interaction with an asymptotic stiff symmetry energy (g0AS) and the Gogny interaction with an asymptotic super-stiff symmetry energy (g0ASS)~\cite{Ono99,Ono2003}. The extracted temperature values are compared to those obtained using a fluctuation thermometer.

\section*{II. Improved MFM Model and Extraction of $a_{sym}/T$}
In Ref.~\cite{Huang2010_1} only yields of isotopes with $I=N-Z=-1, 1$ and 3 were  used to extract the ratio of the symmetry energy coefficient relative to the temperature, $a_{sym}/T$. In this article, an improved method is used to extract the $a_{sym}/T$ values. In the improved method, all available isotope yields are employed. The improvement is made possible by taking into account the mass dependent temperature in the free energy in an iterative manner.

In the framework of MFM, the yield of an isotope with mass $A$ and $I=N-Z$ ($N$ neutrons and $Z$ protons) produced in a multi-fragmentation reaction, can be given as~\cite{Minich1982,Hirsch1984,Bonasera2000,Bonasera2008,Huang2010,Huang2010_1}
\begin{equation}
\begin{split}
Y(I,A) =& Y_{0} \cdot A^{-\tau}exp[\frac{W(I,A)+\mu_{n}N+\mu_{p}Z}{T}\\
&+Nln(\frac{N}{A})+Zln(\frac{Z}{A})].
\end{split}
\label{eq:eq2}
\end{equation}
Using the generalized Weisz$\ddot{a}$cker-Bethe semiclassical mass formula~\cite{Weizsacker1935,Bethe1936}, $W(I,A)$ can be approximated as
\begin{equation}
\begin{split}
W(I,A) =& a_{v}A- a_{s}A^{2/3}- a_{c}\frac{Z(Z-1)}{A^{1/3}}\\
&-a_{sym}\frac{(N-Z)^{2}}{A}
- a_{p}\frac{\delta}{A^{1/2}},\\
\delta =& - \frac{(-1)^{Z}+(-1)^{N}}{2}.
\end{split}
\label{eq:eq3}
\end{equation}
In Eq.\eqref{eq:eq2}, $A^{-\tau}$ and $Nln(N/A)+Zln(Z/A)$ originate from the increases of the entropy and the mixing entropy at the time of the fragment formation, respectively. $\mu_{n}$ ($\mu_{p}$) is the neutron (proton) chemical potential. $\tau$ is the critical exponent. In this work, the value of $\tau=2.3$ is adopted from the previous studies~\cite{Bonasera2008}.
In general coefficients, $a_{v}$, $a_{s}$, $a_{sym}$, $a_{p}$ and the chemical potentials are temperature and density dependent.
In this formulation a constant volume process at an equilibrium, is assumed in the free energy, and therefore the term "symmetry energy" is used throughout this work along Ref.~\cite{Marini2012}. If one assumes a constant pressure at the equilibrium process~\cite{Sobotka2011}, the term "symmetry enthalpy" should be used. Experimentally, whether the equilibrium process takes place at constant pressure or volume can not be determined, and thus we use "symmetry energy" throughout the paper, keeping in mind the ambiguity~\cite{Marini2012}.

In the previous analyses~\cite{Huang2010_1,Huang2010_2,Chen2010,Lin2014,Lin2014_2}, the temperature in Eq.\eqref{eq:eq2} was assumed to be identical to the temperature of the fragmenting source and treated as a constant for all isotopes. However as seen below, this temperature turns out to be (fragment) mass dependent. This mass dependence on the temperature was not recognized in the previous analyses, just because the mass dependence was masked by
the larger error bars. In this improved method, the error bars become small and the mass dependence becomes evident. In order to take into account the mass dependence of the temperature in Eq.\eqref{eq:eq2}, the temperature $T$ is replaced by an apparent temperature $T(A)$. We attribute this mass dependence to the system size effect as discussed in Sec. IV.1. In the improved MFM formulation, therefore, this system size effect is empirically realized by reducing the apparent temperature as $A$ increases as $T(A)=T_0(1-kA)$. $T_0$ is the temperature of the fragmenting source and $k$ is a constant quantifying the mass dependence.

In order to study the density, temperature and symmetry energy in the fragmenting source, the improved MFM of Eq.\eqref{eq:eq2} is utilized to extract the $a_{sym}/T_{0}$ value from the available isotope yields. Since the $a_{sym}/T_{0}$ value in Eqs.\eqref{eq:eq2} and \eqref{eq:eq3} depends on 5 parameters, $a_{v}$, $a_{s}$, $a_{c}$, $a_{p}$ and $\Delta \mu$ (defined by $\Delta \mu = \mu_{n}-\mu_{p}$), the optimization process of these parameters is divided into the following three steps to minimize the ambiguity for each parameter. For a given $k$ value,

\begin{enumerate}
\item{
Optimize $\Delta\mu/T_{0}$ and $a_c/T_{0}$ values from mirror isobars and fix these parameter values.
}
\item{Optimize $a_v/T_{0}$, $a_s/T_{0}$ and $a_p/T_{0}$ values from $N=Z$ isotopes.}
\item{Using extracted parameters in step (1) and step (2), $a_{sym}/T_{0}$ values are extracted from all available isotopes. Comparing the extracted $a_{sym}/T_{0}$ values with three different interactions, the density of the fragmenting source is extracted. Using this density, the value of the symmetry energy coefficient, $a_{sym}$, for each interaction is determined. The temperature is then calculated as the ratio of $a_{sym}$ to $a_{sym}/T_{0}$.
    
}
\end{enumerate}

It is expectable that if the $k$ value is properly selected which means the mass dependence is well considered, a constant $T_{0}$ is obtained. Since the $k$ value is small as seen below, we perform the optimization of the parameter $k$ in an iterative manner in the following analysis, that is, in the first round $k=k_{1}=0$ is set in $T(A)=T_0(1-kA)$ and calculate the temperature as a function of $A$, using steps (1)-(3). From this plot a new $k$ value, $k=k_{1}'$, is extracted from the slope. In the second round, $k=k_{2}=k_{1}+\frac{1}{2}k_{1}'$ is used for the steps (1)-(3) and a new $k$ value, $k=k_{2}'$, is extracted. If $k_{2}'$ is 0 within a given error range, the iteration stops and the $k_{2}$ value is fixed as the mass dependent parameter of the apparent temperature and $T_{0}$ value is determined. Otherwise the iteration continues.

\begin{figure}
\includegraphics[scale=0.45]{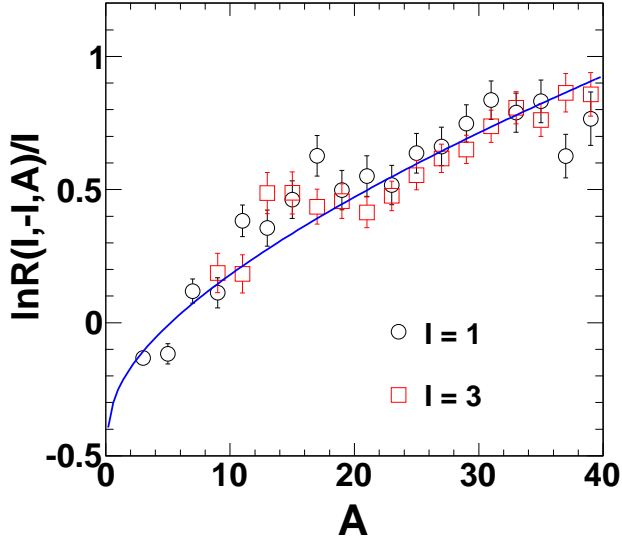}
\caption{\footnotesize (color online)
$ln[R(I,-I,A)]/I$ versus $A$ for $I=1$ and $I=3$ from the events with the g0 interaction. The curve is the fit result of Eq.\eqref{eq:Coulomb} for $k=0$. The extracted values of $\Delta\mu/T_{0}$ and $a_c/T_{0}$ are given in the third and fifth columns of Table~\ref{table:parameters}.
}
\label{fig:fig0}
\end{figure}

Details of steps (1)-(3) are first described below for a given $k$ value.
In the step (1), following Ref.~\cite{Huang2010_1}, the isotope yield ratio between isobars with $I+2$ and $I$, $R(I+2,I,A)$, is utilized, which is
\begin{eqnarray}
&R&(I+2,I,A) = Y(I+2,A)/Y(I,A)   \nonumber\\
&=&  exp\{[\mu_{n}- \mu_{p}+ 2a_{c}(Z-1)/A^{1/3}\nonumber\\
&-&4a_{sym}(I+1)/A-\delta(N+1,Z-1) \nonumber\\
&-& \delta(N,Z)]/[T_{0}(1-kA)]+ \Delta(I+2,I,A)\},
\label{eq:eq_RI2}
\end{eqnarray}
where $Y(I,A)$ is the yield of isotopes with $I$ and $A$, and $\Delta(I+2,I,A)=S_{mix}(I+2,A) - S_{mix}(I,A)$.
When the above equation is applied for a pair of mirror nuclei of odd mass isotopes with $I = -I$ and $I$, the symmetry energy term, pairing term and mixing entropy terms drop out and the following equation is obtained.

\begin{equation}
    \ln[R(I, -I, A)]/I=[\Delta\mu+a_{c}(A-1)/A^{1/3}]/[T_{0}(1-kA)].
\label{eq:Coulomb}
\end{equation}
For all available mirror isobars, $\Delta\mu/T_{0}$ and $a_c/T_{0}$ are optimized in Eq.\eqref{eq:Coulomb}. The $\ln[R(I, -I, A)]$ values and the fit result for $k=0$ is shown in Fig.\ref{fig:fig0}.

\begin{figure}
\includegraphics[scale=0.45]{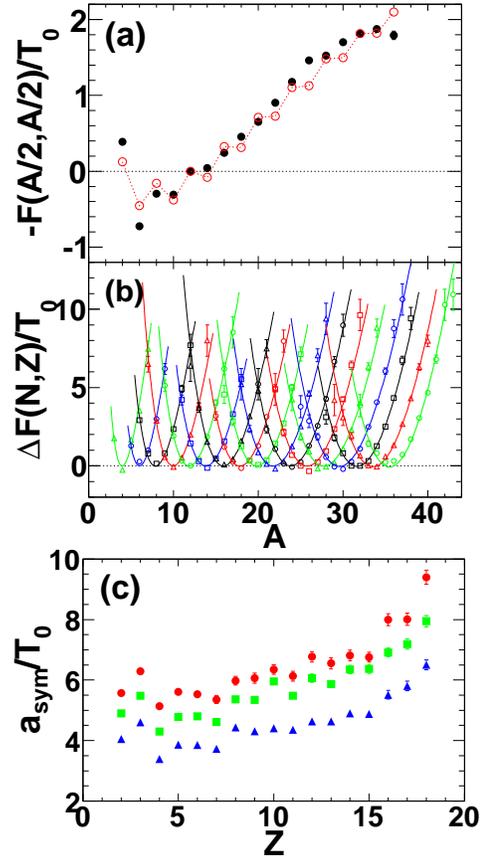}
\caption{\footnotesize (color online)
(a) Calculated ratio of free energy to $T_{0}$ for $N=Z$ isotopes from the AMD events with g0 (solid points). Circles represent the fit using Eq.\eqref{eq:eq5}.
(b) Calculated $\frac{\Delta F(N,Z)}{T_{0}}$ values for the g0 interaction and quadratic fits using Eq.\eqref{eq:eq7} for $Z=2$ to 18. The same symbols are used for isotopes with a given $Z$.
(c) Extracted $a_{sym}/T_{0}$ values from (b) for g0 (dots), g0AS (squares) and g0ASS (triangles). Errors are from the quadratic fits.
All results are from the first round $(k=0)$.
}
\label{fig:fig1}
\end{figure}

In the step (2) we apply Eq.\eqref{eq:eq2} to the isotopes with $N=Z$ with the extracted $\Delta\mu/T_{0}$ and $a_c/T_{0}$ values in the step (1).
For $N = Z = A/2$ isotopes, the ratio of the free energy relative to the temperature can be calculated from Eq.\eqref{eq:eq2} and Eq.\eqref{eq:eq3} without the symmetry energy term as 
\begin{equation}
\begin{split}
-\frac{F(A/2,A/2)}{T_{0}} =& -\frac{F(A/2,A/2)}{T(A)} \cdot (1-kA) \\
=& ln[\frac{Y(A/2,A/2)A^{\tau}}{Y_{0}}] \cdot (1-kA)  \\
=&\frac{\widetilde{a_{v}}}{T_{0}}A- \frac{a_{s}}{T_{0}}A^{2/3}-\frac{a_{c}}{T_{0}}\frac{A(A-2)}{4A^{1/3}}\\
&- \frac{a_{p}}{T_{0}}\frac{\delta}{A^{1/2}}+ A(1-kA)ln(\frac{1}{2}),
\end{split}
\label{eq:eq5}
\end{equation}
where $\widetilde{a_{v}}=a_{v}+\frac{1}{2}(\mu_{n}+\mu_{p})$.
The value of $ln[\frac{Y(A/2,A/2)A^{\tau}}{Y_{0}}]$ on the right of the second equation can be calculated from the simulated or experimental values when the $\tau$ value is fixed.
Non zero values show the deviation of the mass distribution of $N=Z$ isotopes from the power law distribution determined by the critical exponent~\cite{Bonasera2008}. When other $\tau$ values are used, the parameter values change accordingly.
In order to eliminate the constant $Y_{0}$, all isotope yields are normalized to the yield of $^{12}C$~\cite{Bonasera2008,Huang2010,Huang2010_1}. For the first round with $k=0$, the renormalized values of $-\frac{F(A/2,A/2)}{T_{0}}$ from the AMD events with the g0 interaction are plotted as a function of the isotope mass $A$ using solid points in Fig.~\ref{fig:fig1}(a). The values of $\widetilde{a_{v}}/T_{0}$, $a_{s}/T_{0}$ and $a_{p}/T_{0}$ are used as free parameters to fit the given $-\frac{F(A/2,A/2)}{T_{0}}$ values, employing Eq.\eqref{eq:eq5}. A typical search result is shown by open circles in Fig.~\ref{fig:fig1}(a) for the case of the g0 interaction. Similar quality results are obtained for the events generated using the g0AS and g0ASS interactions.
One should note that the value of $a_p/T_{0}$ make a small contribution and the contribution is evident as a staggering in the $-F(A/2,A/2)/T_{0}$ vs $A$ plot. Therefore the essential free parameters in this step are $\widetilde{a_{v}}/T_{0}$ and $a_{s}/T_{0}$.
The extracted parameter values are summarized in Table~\ref{table:parameters} for the first round $(k=0)$ and the final round $(k=0.007)$.

\begin{table}[ht]
\caption{$a/T_{0}$ and $\Delta \mu/T_{0}$ for the first round ($k=0$) and the final round($k=0.007$)} 
\centering 
\begin{tabular}{c c c c c c} 
\hline\hline 
      & $\widetilde{a_{v}}/T_{0}$ & $a_{s}/T_{0}$ & $a_{c}/T_{0}$ & $a_{p}/T_{0}$  & $\Delta \mu /T_{0} ^{~a}$ \\ [.5ex] 
\hline 
k=0.0 & & & & &  \\
\hline 
g0    &    1.77       &    2.74       & $1.04\times 10^{-1}$ & $4.27\times 10^{-1}$ & $-2.54\times 10^{-1}$\\
g0AS  &    1.76       &    2.66       & $1.08\times 10^{-1}$ & $6.61\times 10^{-1}$ & $-2.72\times 10^{-1}$\\
g0ASS &    1.77       &    2.72       & $1.03\times 10^{-1}$ & $7.78\times 10^{-1}$ & $-2.52\times 10^{-1}$\\
\hline 
k=0.007 & & & & &  \\
\hline 
g0    &    1.55       &    2.29       & $8.05\times 10^{-2}$ & $4.16\times 10^{-1}$ & $-1.89\times 10^{-1}$\\
g0AS  &    1.53       &    2.20       & $8.30\times 10^{-2}$ & $6.51\times 10^{-1}$ & $-2.02\times 10^{-1}$\\
g0ASS &    1.55       &    2.29       & $8.00\times 10^{-2}$ & $7.66\times 10^{-1}$ & $-1.86\times 10^{-1}$\\
\hline 
\end{tabular}
\footnote {$\Delta \mu /T_{0}$ values are taken from the step (1).}
\label{table:parameters} 
\end{table}

In step (3) Eq.\eqref{eq:eq2} is applied to yields of all isotopes with $N=Z$ and $N \ne Z$. From Eq.\eqref{eq:eq2}, $a_{sym}/T_{0}$ and $\Delta \mu / T_{0} $ values can be related to the modified free energy, $\frac{\Delta F(N,Z)}{T_{0}}$ as

\begin{equation}
\begin{split}
\frac{\Delta F(N,Z)}{T_{0}}=& \frac{a_{sym}}{T_{0}}\frac{(N-Z)^{2}}{A}-\frac{\Delta \mu}{2T_{0}}(N-Z),\\
\end{split}
\label{eq:eq7}
\end{equation}
where $\frac{\Delta F(N,Z)}{T_{0}}$ is the free energy relative to the temperature,  $\frac{F(N,Z)}{T_{0}}$, subtracted by the calculated contributions of the volume, surface, Coulomb and pairing terms, using the parameters in Table~\ref{table:parameters}. Resultant $\frac{\Delta F(N,Z)}{T_{0}}$ values are shown by symbols in Fig.~\ref{fig:fig1}(b). They exhibit quadratic relationships with minimum values close to zero.
The minimum values are at or near $N=Z$ isotopes and therefore reflect approximately the difference between the data and fits point in Fig.~\ref{fig:fig1}(a).
In this step, the $a_{sym}/T_{0}$ and the $\Delta \mu / T_{0}$ values are optimized. Since the $\Delta \mu / T_{0}$ values are extracted from the step(1), the optimization is made around the values in the fifth column of Table~\ref{table:parameters} in a small margin. The $a_{sym}/T_{0}$ values are extracted from the quadratic curvature of the isotope distribution for each given $Z$ and plotted in Fig.~\ref{fig:fig1}(c) separately for the g0, g0AS and g0ASS interactions.
As one can see for the first round with $k=0$, the extracted $a_{sym}/T_{0}$ values increase as $Z$ increases in all cases, and they more or less parallel each other.

\section*{III. Self-consistent Determination of Density and Temperature}

\begin{figure}
\includegraphics[scale=0.45]{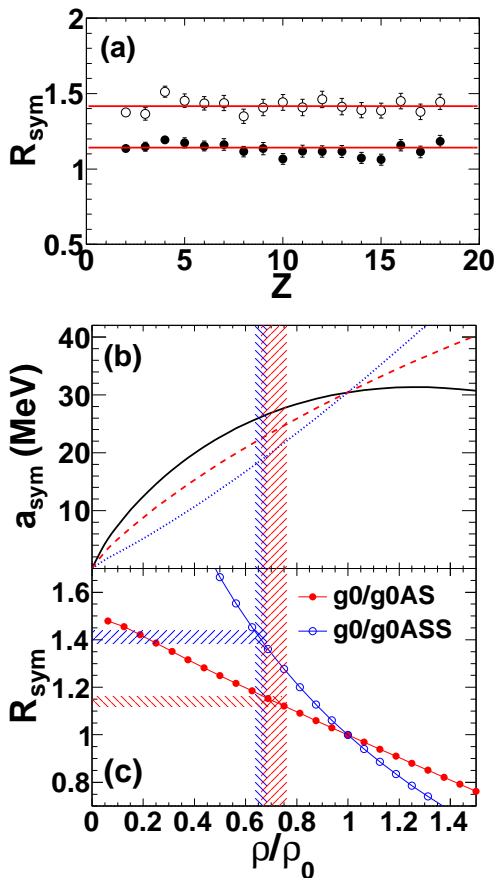}
\caption{\footnotesize (color online)
(a) The ratios of the $a_{sym}/T_{0}$ values shown in Fig.~\ref{fig:fig1}(c), dots for g0/g0AS and circles for g0/g0ASS. (b) Symmetry energy coefficient vs density used in the AMD simulations. Solid curve(g0), dashed (g0AS) and dotted (g0ASS) (c) The ratio of the symmetry energy coefficient in (b). The shaded horizontal lines indicate the ratios extracted in (a) and the vertical shaded area shows the density region corresponding to these ratios. Two different shadings are used for the two ratio values. All results are from the first round.
}
\label{fig:fig2}
\end{figure}

In order to determine the density and temperature at the time of the fragment formation, the parallel behavior of the observed $a_{sym}/T_{0}$ values in Fig.~\ref{fig:fig1}(c) is utilized.
As suggested in Ref.~\cite{Ono2003}, the observed differences are attributed to the difference of the symmetry energy at the density at the time of the fragment formation. The ratios between g0/g0AS and g0/g0ASS of $a_{sym}/T_{0}$ for the first round are shown in Fig.~\ref{fig:fig2}(a). The ratios show flat distributions as a function of $Z$ for both cases. The extracted average ratio values are shown by lines in the figure and the values are given in the first column of Table~\ref{table:data_results}. In Fig.~\ref{fig:fig2}(b) the symmetry energy coefficient is plotted as a function of the density for the three interactions used in the calculations and in Fig.~\ref{fig:fig2}(c) their ratios, $R_{sym} = a_{sym}(g0)/a_{sym}(g0AS)$ and $R_{sym} = a_{sym}(g0)/a_{sym}(g0ASS)$, are plotted. Using the ratio values determined from Fig.~\ref{fig:fig2}(a) and the density dependence of the $R_{sym}$ values in Fig.~\ref{fig:fig2}(c), the implied densities of the fragmenting sources are indicated by the shaded vertical areas shown in Fig.~\ref{fig:fig2}(c). The extracted density values for each case are given in the second column of Table~\ref{table:data_results}. Assuming that the nucleon density should be same for the three different interactions used, the nucleon density of the fragmenting source is determined from the overlap of the extracted values. This assumption is reasonable for the central collisions because the nucleon density is mainly determined by the stiffness of the EOS and not by the density dependence of the symmetry energy term. From the overlapped density area in Figs.~\ref{fig:fig2}(c), $\rho/\rho_0 = 0.67 \pm 0.02$ is extracted as the density at the time of the fragment formation. Using this density value, the corresponding symmetry energy values at that density are extracted for the three different interactions from Fig.~\ref{fig:fig2}(b). They are given in the third column of Table~\ref{table:data_results}.

\begin{table}[ht]
\caption{symmetry energy and $\rho/\rho_{0}$ from $k=0.007$} 
\centering 
\begin{tabular}{c c c c c} 
\hline\hline 
k&int& $R_{sym}$ & $\rho/\rho_{0}$ & $a_{sym}$ (MeV)\\ [0.5ex] 
\hline 
k=0.0&g0    &               &               & 26.6$\pm$0.3 \\
&g0/g0AS       & 1.14$\pm$0.02 & 0.71$\pm$0.05 &              \\
&g0AS  &               &               & 23.7$\pm$1.3 \\
&g0/g0ASS      & 1.41$\pm$0.03 & 0.66$\pm$0.02 &              \\
&g0ASS &               &               & 18.7$\pm$0.7 \\
\hline 
k=0.007&g0    &               &               & 26.8$\pm$0.2 \\
&g0/g0AS      & 1.14$\pm$0.02 & 0.72$\pm$0.05 &              \\
&g0AS  &               &               & 24.0$\pm$1.3 \\
&g0/g0ASS      & 1.40$\pm$0.03 & 0.66$\pm$0.02 &              \\
&g0ASS &               &               & 18.7$\pm$0.7 \\
\hline 
\end{tabular}
\label{table:data_results} 
\end{table}

\begin{figure}
\includegraphics[scale=0.45]{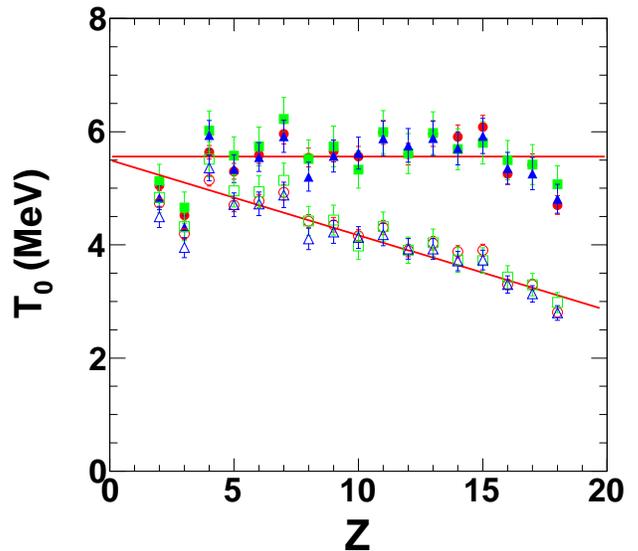}
\caption{\footnotesize (color online)
Extracted $T_0$ values as a function of $Z$. Open symbols are for the first round ($k=0$) and
closed symbols are for the final round ($k=0.007$). Different symbols represent results with $g0$(circles), $g0AS$(squares) and $g0ASS$(triangles) interactions.
}
\label{fig:fig3}
\end{figure}

Once the symmetry energy value is determined for a given interaction, the temperature, $T_0$, can be calculated as $T_{0} = a_{sym}$ / ($a_{sym}/T_{0})$. The extracted $T_0$ values are shown as a function of $Z$ by open symbols for the first round in Fig.~\ref{fig:fig3}.
The larger errors of $T_{0}$, comparing to those in Fig.~\ref{fig:fig1}(c), originate from the errors of $a_{sym}$ and $a_{sym}/T_{0}$ extracted for each interaction which are shown in the third column of Table~\ref{table:data_results} and Fig.~\ref{fig:fig1}(c), respectively. The temperature values extracted from the three different interactions agree with each other very well and show a monotonic decrease as $Z$ increases from $\sim 5 $ MeV at $Z =4$ to $\sim 3 $ MeV at $Z = 18$. From this slope the extracted temperature as function of $A$, $T_{0} =5.5( 1 - 0.012A)$, is determined for the first round, assuming $A \sim 2Z$.

\begin{figure}
\includegraphics[scale=0.45]{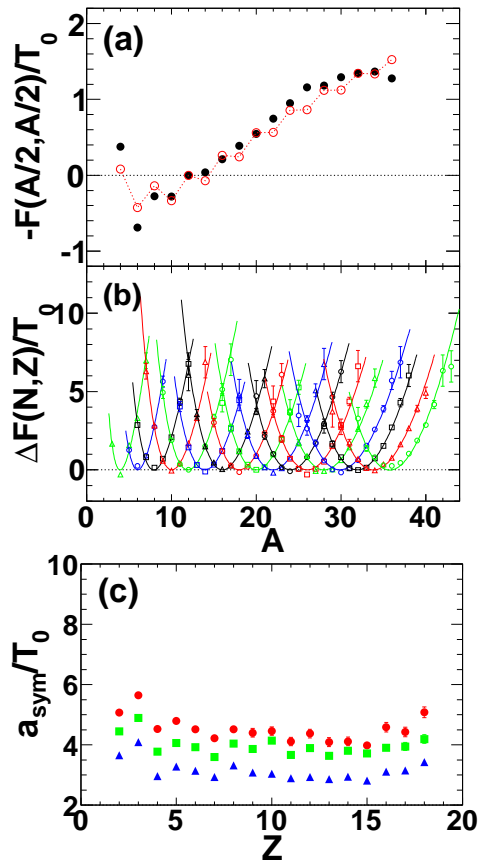}
\caption{\footnotesize (color online)
Same plots as Fig.\ref{fig:fig1}, but for the final round ($k=0.007$).  See also the figure caption of Fig.\ref{fig:fig1}.
}
\label{fig:fig4}
\end{figure}

The iteration are repeated four times in this work.
The same plots as Fig.\ref{fig:fig1}, but with the $k$ value for the fourth (final) round, $k=0.007$, are shown in Fig.\ref{fig:fig4} and the extracted parameters are also given in Table~\ref{table:parameters}. A very similar quality of results to those of the first round with $k=0$ are obtained, even though the optimized parameter values are quite different between those of the first round $(k=0)$ and of the fourth round $(k=0.007)$.
The extracted $a_{sym}/T_{0}$ values parallel each other and show a rather flat distribution as a function of $Z$ for $Z$ up to 15 in Fig.~\ref{fig:fig4}(c).
As seen in Fig.~\ref{fig:fig3}, in which the extracted $T_0$ values are shown by closed symbols as a function of $Z$ for the fourth round with $k=0.007$, the extracted $T_0$ values are consistent with 5.5 MeV within the error bars.
Since $T_{0}$ values show a flat distribution as a function of $Z$, the iteration is stopped at this round and $T_{0}=5.5\pm 0.2$ MeV is taken as the temperature of the emitting source.

The extracted parameter values of $R_{sym}$, density and symmetry energy values for the fourth round are very similar to those of the first round as shown in Table~\ref{table:data_results}.
The values and errors of these parameters are essentially determined by the ratios and their errors of the $a_{sym}/T_{0}$ values between different interactions as discussed in Fig.\ref{fig:fig2}(a).
These ratio values are stable between the first and fourth rounds, even though the optimized parameters values in Table~\ref{table:parameters} are quite different between these rounds.
The extracted value of the  mass dependent factor $k$ has some errors, but the above fact ensures that the extracted density, temperature and symmetry values and their errors in Table~\ref{table:data_results} are rather stable independent of the choice of the $k$ values within its error bar, when the parameters in Table~\ref{table:parameters} are optimized for the given $k$ value.

The decreasing trend of the temperature as $A$ increases is often observed in heavy ion collisions and normally attributed to variations in impact parameter~\cite{Trautmann07}. The heavier fragments tend to be produced in more peripheral collisions and therefore show lower temperature. However in this study
all events analyzed are generated in the same class of events, central collisions with $b=0$ fm. Therefore we attribute the decreasing trend may originate to the different fragment formation processes rather than to the centrality of the events.

\section*{IV. Discussion}

\section*{IV.1 Temperature and Fluctuation thermometer}

The temperature at the time of the fragment formation has been studied, using different thermometers~\cite{Guo2013} as mentioned earlier.
Here we will investigate the applicability of the fluctuation thermometer
using different isotopes, which gives the temperature values as a function of fragment mass A.

For a given variable $X$, the fluctuation is given as $\sigma^2(X) = \langle X^{2} \rangle - \langle X \rangle^{2}$.
In our present work, focusing on momentum fluctuations,  $X = Q_{xy}=P_{x}^2 - P_{y}^2$ is used, following Ref.~\cite{Wuenschel2010}.
Different relationships have been proposed to derive the temperature from observed momentum fluctuations.
In Refs.~\cite{Zheng2011,Zheng2012}, Zheng et al. reported that there may be significant effects on the fluctuation temperature resulting from the quantum nature of the particles examined. When the temperature is determined from the fermions, such as n, p, t etc., the momentum distribution is expected to follow the Fermi-Dirac distribution ~\cite{Zheng2011}. On the other hand, when bosons, such as d or $\alpha$ particles, are used, the Bose-Einstein distribution is expected to be appropriate~\cite{Zheng2012}.
In this paper the fluctuation temperature is calculated from IMF momentum distributions including those in the excited states from the primary fragments of AMD simulations without the afterburner. Even for a given IMF isotope, therefore, fermions and bosons are mixed and we assume the classical momentum distribution for these IMFs in the following analysis.

\begin{figure}
\includegraphics[scale=0.35]{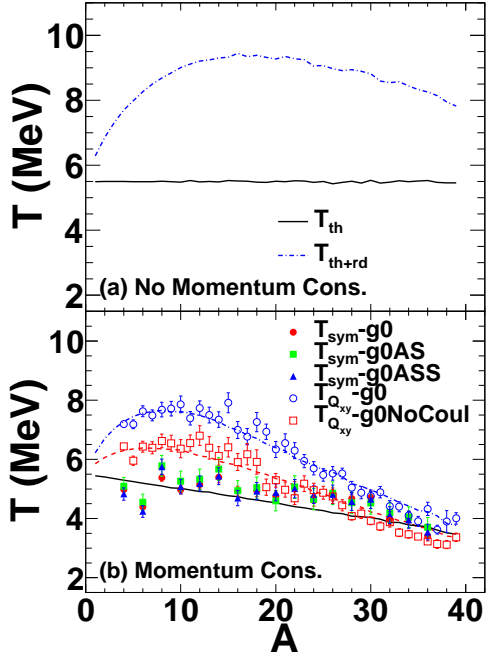}
\caption{\footnotesize (color online)
(a) Temperatures from the Monte-Carlo model calculation. The solid curve represents the fluctuation temperature values for a thermally equilibrated source using $T_0=5.5$ MeV. The dashed curve represents those corresponding to the source with additional radial flow with $f_r=0.022$ c/fm. No momentum conservation is applied. (b) Temperature values extracted in different methods. Closed symbols (dots, squares, triangles) represent the apparent temperature values from the self-consistent symmetry energy analysis, using the AMD events from the final round with $k=0.007$. Open circles represent the fluctuation temperature values, $T_{Q_{xy}}$. Open squares represent those from the AMD events with no Coulomb potential. Solid and dashed curves are those similar to (a) but under the condition of the momentum conservation.
}
\label{fig:fig5}
\end{figure}

Under the assumption of the Maxwell-Boltzmann distribution, the momentum distribution is,
\begin{equation}
\begin{split}
f(\overrightarrow{P})=\frac{1}{(2m\pi T)^{3/2}}exp(-\frac{\overrightarrow{P}^{2}}{2mT}),
\end{split}
\label{eq:eq_P}
\end{equation}
and the fluctuation of $Q_{xy}$, $\sigma^2(Q_{xy})$, can be given as
\begin{equation}
\begin{split}
\sigma^2(Q_{xy})= \int (P_{x}^{2}-P_{y}^{2})^{2} f(\overrightarrow{P})d\overrightarrow{P}.    \\
\end{split}
\label{eq:eq9}
\end{equation}
From above one can show that $\sigma^2(Q_{xy}) = 4m^2T^2$ for a given particle with mass $m$, and therefore the fluctuation temperature $T_{Q_{xy}}=\sqrt{\sigma^2(Q_{xy})}/2m$ is obtained.
The fluctuation temperature values calculated from the AMD events with the g0 interaction are represented by open circles in Fig.~\ref{fig:fig5}(b). Very similar results of the fluctuation temperature values are obtained for the other two interactions. As seen in the figure, over the entire range of isotope mass, the fluctuation temperature values are quite different from the $T_{sym}=T(A)$ values in the previous section. $T_{Q_{xy}}$ shows a broad peak $T_{Q_{xy}}\sim8$ MeV around $A \sim 12$, and then a monotonic  decrease  as $A$ increases. For comparison the fluctuation temperature values from the AMD events without the Coulomb interaction are also calculated and shown by open squares. They show a similar trend to those with the Coulomb interaction included, but about 1.0 to 1.5 MeV lower temperature in the entire mass range.

In order to understand the behavior of the extracted fluctuation temperature, simple Monte-Carlo calculations are performed.
In this model, a percolation model is utilized to generate fragments from
80 lattice points ($4 \times 4 \times 5$)~\cite{Staufer1979}.
No charge and internal excitation energy are considered in the percolation model.
The motion of particles was assigned as the sum of the thermal motion and radial expansion.
One can use a more sophisticated method to generate the fragments, such as those used in statistical multifragmentation model (SMM)~\cite{Bondorf85} or microcanonical Metropolitan Monte Carlo model (MMMC)~\cite{Gross90}, but here we just use a simpler model available for our purpose, because we are only concerned with the distribution of thermal and collective motions assigned to these fragments.
In the central AMD collisions ($b=0$ fm), collective contributions to the particle energy occur only in the radial direction. Both radial flow and Coulomb repulsion contribute to the radial expansion motion. The Coulomb acceleration is treated as a part of the radial flow and the radial velocity is defined as $\overrightarrow{v^{rd}} = f_r\overrightarrow{r}$, where $f_r$ is the flow parameter and $\overrightarrow{r}$ is the coordinate vector. Fragments generated by the percolation model are distributed uniformly inside a sphere of radius $r_{sys}$ under the condition of $r \le r_{sys} - r_A$, where $r_A$ is the radius of the fragment with mass A. They are calculated as
$r_{sym}=1.2\times[80/(\rho/\rho_{0})]^{1/3}$ and $r_A=1.2\times [A/(\rho/\rho_{0})]^{1/3}$, where $\rho/\rho_{0}=0.67$ is adopted from the above analysis.
This condition allows some overlap of two spheres but for the purpose of the restriction of the $r$ distribution, this condition is good enough to see the global effect of the fluctuation temperature.
Under a thermal equilibrium condition, the thermal velocity $v^{th}_{i}$, where $i=x,y,z$, is given by a Maxwell-Boltzmann velocity distribution as
\begin{equation}
\begin{split}
f(v^{th}_{i}) = \sqrt{\frac{A}{2\pi T_{0}}} exp[-\frac{(v_{i}^{th})^{2}}{2 \cdot (T_0/A)}],\\
\end{split}
\label{eq:eq_th}
\end{equation}
where $T_0$ is the input parameter in the model. The total velocity $\overrightarrow{v}$ is the sum of these two velocity vectors, $\overrightarrow{v}=\overrightarrow{v^{th}}+\overrightarrow{v^{rd}}$.
For a given parameter set of $f_r$ and $T_0$ values, which are determined below, the temperature is evaluated using the fluctuation thermometer and the extracted temperature values are shown in Fig.~\ref{fig:fig5}(a) for the thermal motion only, $T_{th}$ (solid curve), and for the sum of the thermal motion and radial motion, $T_{th+rd}$ (dashed curve). As expected, the temperature values for the thermal motion show a constant temperature. The temperature with the radial flow does not increase linearly as $A$ increases because of the condition on the radial distance of the fragments $r \le r_{sys} - r_A$. In order to apply this model to the reaction events, momentum conservation, $\sum_j{m_j\overrightarrow{v(j)}=0}; (j$ for all fragments in one event), has to be considered. In the actual application, among more than a hundred million events used in Fig.~\ref{fig:fig5}(a), only those with $|\sum_j{m_j\overrightarrow{v(j)}| \le 100}$ MeV/c are selected as an approximation of $\sum_j{m_j\overrightarrow{v(j)}=0}$. When the momentum conservation of the system is required, the momenta of the larger fragments tend to be limited in smaller side relative to those of $\overrightarrow{v}=\overrightarrow{v^{th}}+\overrightarrow{v^{rd}}$ used in Fig.~\ref{fig:fig5}(a). The results are shown by solid and dashed lines in Fig.~\ref{fig:fig5}(b). The parameters $T_0$ and $f_r$ are adjusted to reproduce the fluctuation temperature, $T_{Q_{xy}}$ extracted from the AMD events in Fig.~\ref{fig:fig5}(b). As seen in the figure, $T_{th}$ and $T_{th+rd}$ show additional significant decreasing trends for heavier fragments, compared to those shown in Fig.~\ref{fig:fig3}(a). Using the parameter sets of $T_0=5.5$ MeV and $f_r=0.022$ c/fm ($f_r=0.016$ c/fm without Coulomb), the $T_{Q_{xy}}$ values from both AMD calculations with and without Coulomb are reproduced reasonably well by $T_{th+rd}$. At the same time, $T_{th}$, the radial flow corrected temperature, reproduces well the temperature values of $T_{sym}$ from the self-consistent symmetry energy analysis of the AMD events.

\section*{IV.2 Radial flow}

\begin{figure}
\includegraphics[scale=0.4]{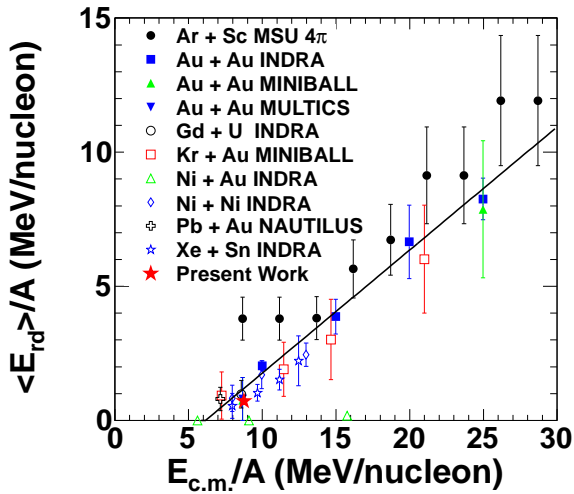}
\caption{\footnotesize (color online)
Radial flow energy vs the available energy. The experimental data are taken from Ref.~\cite{Borderie08}.
}
\label{fig:fig6}
\end{figure}

From the above  Monte-Carlo simulation, the average radial flow energy, excluding the Coulomb contribution, is evaluated as $\langle E_{rd}\rangle = 0.72$ MeV/nucleon
from Fig.~\ref{fig:fig5}(b). The average Coulomb repulsive energy is evaluated as $\langle E_{Coul}\rangle = 0.61$ MeV/nucleon. Since the radial flow energy is not modified by the secondary decay cooling process, the extracted radial flow energy in the above analysis can be compared to the existing experimental data as shown in Fig.~\ref{fig:fig6}. Most of the existing data are distributed along the line $\langle E_{rd}\rangle = 0.46 (E_{cm}/A - 6.1)$ and, as seen in the figure, the extracted radial flow energy is consistent with that trend.

\section*{IV.3 Volume and surface contribution of Symmetry energy}
\begin{figure}
\includegraphics[scale=0.4]{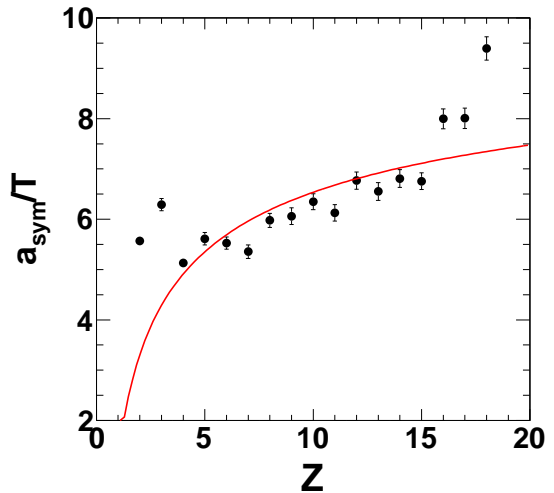}
\caption{\footnotesize (color online)
Fit results of $a_{sym}/T$ from Fig.\ref{fig:fig1}(c) of the first round (k=0) with a function, $a_{sym}/T = a^{(V )}_{sym}/T(1-k_{S/V} A^{-1/3})$. $a^{(V )}_{sym}/T = 11.1$ and $k_{S/V}=1.2$ is used.
}
\label{fig:fig7}
\end{figure}

The increase of $a_{sym} / T$ seen in Fig.\ref{fig:fig1}(c) can be attributed to the surface contribution of the symmetry energy in finite nuclei, assuming a constant temperature. In the Weisz$\ddot{a}$cker-Bethe semiclassical mass formula used in Eq.\eqref{eq:eq3}, the coefficient, $a_{sym}/T$, is independent of the nuclear size, which originates from the volume nature of the symmetry energy. However advanced mass formulas have introduced size dependence on  $a_{sym} / T$ reflecting the surface effect~\cite{Mayer1966,Danielewicz2003}.
Following the formulation of Danielewicz et al.~\cite{Danielewicz2003,Danielewicz2009}, the symmetry coefficient can be expressed as $a_{sym}/T = a^{(V )}_{sym}/T(1-k_{S/V} A^{-1/3})$ where $k_{S/V} = a^{(S)}_{sym}/a^{(V )}_{sym}$ and the suffixes V,S denote volume and surface contributions, respectively. One should note that the $k_{S/V}$ value is independent of the temperature $T$ of the emitting source.
$k_{S/V} = 1.67 \sim 1.84 $ has been obtained from ground state nuclei~\cite{Jiang2012}. When we apply this formula to fit the results in Fig.~\ref{fig:fig1}(c) for the g0 interaction, one can get $a^{(V )}_{sym}/T = 11.1$ and $k_{S/V} = 1.2$. The fit result is shown in Fig.\ref{fig:fig7}. If we assume $a_{sym} = a^{(V )}_{sym}$ and use the value $a_{sym}=26.8$ MeV from Table~\ref{table:data_results}, we get $T \sim 2.4$ MeV, which is much lower than that extracted above from the fluctuation thermometer.
The extracted temperature independent $k_{S/V}$ value, $k_{S/V}=1.1$ is also significantly smaller than values extracted from the ground state nuclei. Since the mass dependence of the temperature extracted above is consistent to the fluctuation temperature after the radial flow correction, we conclude that the mass dependence of the $a_{sym}/T$ values in Fig.~\ref{fig:fig1}(c) originates from the mass dependent apparent temperature and that the surface contribution of the symmetry energy is small, if any. This conclusion is also consistent with  our recent results of $^{64}$Zn + $^{112}$Sn at 40 MeV/nucleon, a slightly larger system, when the improved MFM method is applied to the reconstructed isotope yields~\cite{Liu2014}. In the analysis a slightly smaller mass dependence of $a_{sym}/T$ values is observed. If the mass dependence originates from the surface contribution, the mass dependence should remain same. On the other hand this smaller mass dependence is consistent with the Monte-Carlo analysis in the previous section for the system size of $a_{sys} \sim 180$.

\section*{V Conclusions and Summary}

The following conclusions are drawn from the above analysis from central collision events of AMD for $^{40}$Ca + $^{40}$Ca at 35 MeV/nucleon:
\begin{enumerate}
\item {
The density of the fragmenting source is evaluated as $\rho/\rho_0=0.67 \pm 0.02$, using the extracted $a_{sym}/T_{0}$ values, based on the improved MFM model and the comparisons between the results from different density dependent symmetry energy terms.
}
\item {The apparent temperature values show a monotonic decrease as the fragment $Z$ decreases from $T\sim 5 $ MeV at $Z=4$ to $T\sim 3$ MeV at $Z=18$.
    This gives the mass dependent apparent temperature $T(A)=5.5(1-kA), k=0.007$.
    }
\item {The fluctuation thermometer, uncorrected for flow effects,  gives larger temperature values and a different shape compared to the apparent temperatures extracted from above. They show a structure, i.e., a broad peak at $ A \sim 12$ and a monotonic decrease for larger $A$. Those from the AMD events without the Coulomb interaction show a similar trend but exhibit about $1\sim 1.5$ MeV lower values.
    }
\item {After the correction for the radial flow requiring momentum conservation in a Monte-Carlo model, the behavior of the fluctuation thermometer values are well understood as the combination of the thermal and radial collective motions. The bell shape structure of the temperature values is well reproduced by the model. The mass dependence of the apparent temperature observed from the self-consistent symmetry energy analysis originates from the system size effect through momentum conservation. This indicates that the fluctuation thermometer can be a reasonable probe to measure the temperature of the fragmenting source in a multi-fragmentation process, if one can properly correct the contribution of the collective motion.
    The evaluated expansion energy from the fluctuation thermometer is consistent with existing experimental values.
    }
\item {Surface contribution of the symmetry energy to the $a_{sym}/T$ values is examined.  We conclude that the surface contribution in $a_{sym}$ is small, if any.
    }
\end{enumerate}

Summarizing, an improved method is proposed for the extraction of the ratio of the symmetry energy coefficient relative to the temperature, $a_{sym} / T(A)$, taking into account the mass dependence of the apparent temperature, based on the MFM model. This method is applied for the central collisions of the AMD events generated  for $^{40}Ca + ^{40}Ca$ at 35 MeV/nucleon.  Gogny interactions, g0, g0AS and g0ASS, with three different density dependencies of the symmetry energy are employed. As a function of IMF charge $Z$, the ratios of the extracted $a_{sym} / T_{0}$ values from the different interactions are essentially constant and reflect the differences of the symmetry energy at the density at the time of the fragment formation. Using this correlation, $\rho/\rho_0 = 0.67 \pm 0.02$ is evaluated as the density at the time of fragmenting source and the symmetry energy value at that density are extracted for each interaction. The temperature values are then determined.  The extracted temperatures show a monotonic decrease as the fragment $Z$ increases, changing from 5 MeV to 3 MeV when $Z$ increases from 4 to 18. The extracted temperature values are compared to those of the fluctuation thermometer, assuming $A\sim 2Z$. The fluctuation thermometer shows significantly higher temperature values and a different trend as a function of fragment mass. After the correction for the radial flow the trends and the differences are well understood as resulting from the combination of the thermal and radial collective motions under  momentum conservation. The mass dependence of the apparent temperature observed from the improved method originates from the system size effect under this momentum conservation. This indicates that the fluctuation thermometer can provide a reasonable probe for the temperature of the multi-fragmentation source if the contribution of collective flows is properly corrected.

\section*{ACKNOWLEDGMENTS}
X.Liu and W. Lin thank Prof. J. B. Natowitz for his hospitality during the stay in the Cyclotron Institute, TAMU. This work is supported by the National Natural Science Foundation of China (Grants No. 11075189) and 100 Persons Project (0910020BR0, Y010110BR0) and ADS project 302 (Y103010ADS) of the Chinese Academy of Sciences, the U.S. Department of Energy under Grant No. DE-FG03-93ER40773 and the Robert A. Welch Foundation under Grant A0330.



\end{document}